# Towards Public Key Infrastructure less authentication in Session Initiation Protocol

Abdullah Al Hasib[1], Abdullah Azfar[2] and Md. Sarwar Morshed[3]

[1,3] Department of Computer Science and Information Technology, Islamic University of Technology
Boardbazar, Bangladesh

[2] School of Information and Communication Technology, Royal Institute of Technology (KTH)
Stockholm, Sweden

## Abstract

The Session Initiation Protocol (SIP) has become the most predominant protocol for Voice over Internet Protocol (VoIP) signaling. Security of SIP is an important consideration for VoIP communication as the traffic is transmitted over the insecure IP network. And the authentication process in SIP ranges from pre-shared secret based solutions to Public Key Infrastructure (PKI) based solution. However, due to the limitations in PKI based solutions, some PKI less authentications mechanisms are proposed. This paper aims to present an overview of different authentication methods used in or together with SIP. We start by highlighting the security issues in SIP in the context of VoIP communication. Then we illustrate the current activities regarding the SIP authentication mechanisms including the recent developments in the research community and standardization efforts within the Internet Engineering Task Force (IETF). Finally we analyze the security aspects of these approaches.

***Keywords:*** *SIP, VoIP, Authentication, Key exchange, Public Key Infrastructure.*

## 1. Introduction

The SIP [1] is an application-layer protocol for managing sessions standardized by the IETF. SIP is a protocol generic enough for being used not only for Internet-Telephony, but for any application that requires the management of sessions, such as multimedia distribution, instant messaging [2], online gaming [3] etc. In the specific case of VoIP, SIP appears as an alternative to ITU-T's (ITU Telecommunication Standardization Sector) H.323 [4] standard and proprietary standards such as the standard used by Skype. SIP is used as the signaling protocol whereas Real-Time Transport Protocol (RTP) is used to provide the multimedia channel. SIP gained greater attention being chosen as the signaling protocol for the 3GPP (Third-Generation Partnership Project) IP Multimedia System [5].

The SIP message flow can be described with the trapezoid illustrated in Fig. 1, where the SIP User Agent client tries to establish a media session with the SIP User Agent Server of another terminal (it is necessary to the SIP terminal to have both agents in order to initiate sessions and respond to requests).

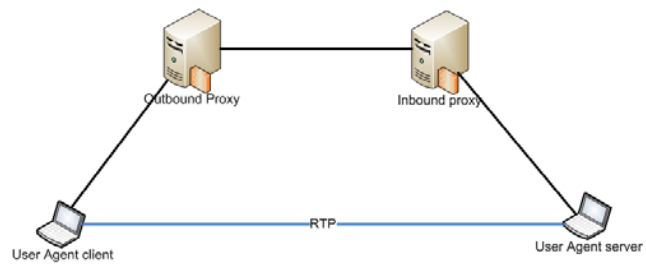

Fig. 1 SIP Trapezoid.

Initially, the client possesses only the Uniform Resource Identifier (URI) of the other user agent. Afterwards, the client forwards the session invitation to the Outbound Proxy Server, which will consult a Domain Name System (DNS) server to find the Inbound Proxy Server for the destination User Agent. Once the Outbound Proxy forwards the invitation to the Inbound Proxy, this last one will look up for the location of the destination agent in the Location Server and then forward the invitation to its final destination. Until the session is established, the messages follow the path: User Agent Client - Outbound Proxy - Inbound Proxy – User Agent Server. Afterward, the media session is transmitted directly between the User Agents through RTP instead of using SIP. A simple message session set-up in SIP can be described as a 3-way handshake with an invitation message from the caller, an OK answer from the destination and, an ACK from the initiator. On the other hand, for the session termination, one BYE message is enough. Several other scenarios are presented [6], and several method extensions have been proposed to the protocol, such as: the SIP info method and the extension for instant messaging [2].

When it comes to the security of SIP, several aspects must be taken into consideration. First of all, SIP carries data that can be considered sensitive (such as the parties URIs, which could consist of their phone numbers, IP addresses) and it indirectly generates a traffic pattern that can potentially give some information about the call. Besides confidentiality, there is the need to assure the identity of the communicating parties and the integrity of the messages on the session management, so an attacker can't alter the session or masquerade a user. There are also the issues of attacks against the protocol itself such as Denial of Service or SPAM over Telephony IP (SPIT). At last, SIP is also one of the channels to exchange or agree in the key material used to protect the media traffic.

The purpose of this paper is to describe different authentication methods that are behind the SIP and analyze the security aspects of the available mechanisms.

The remainder of this paper is organized as follows. In the section 2, we present and discuss the SIP Authentication Mechanisms. Then, at section 3, we elaborate some PKI less alternatives which encompass authentication and key agreement at the same time. Later, we present an analysis of different the authentication mechanisms in section 4. Finally, the paper is ended with the conclusion at section 5.

## 2. SIP Authentication Mechanisms

Authentication process is necessary to ensure that the communication is going to take place only among the legitimate users. In SIP, proper authentication is necessary while the participant wants to register or setup, modify or terminate as session. Different authentication mechanisms work at different layers. Table 1 presents the SIP authentication mechanism at different layers.

### 2.1 SIP Digest Authentication

SIP Digest Authentication mechanism [1] is originally based on HTTP Digest mechanism [7]. This is a challenge-response paradigm which is used for client-to-client or client-to-proxy authentication. In this scheme, the recipient can challenge the identity of the sender using a nonce value. In response to this challenge, the sender sends a message digest calculated using this nonce, username, shared secret and some other optional parameters.

Table 1: SIP Authentication mechanisms at different layers

|  | HTTP Basic Authentication | SIP Digest Authentication |
|---|---|---|
| Application Layer Mechanism |  | Extended SIP Digest Authentication |
|  | S/MIME | SIP AIB |
|  | ID-Based Authentication | Certificate-less Authentication |
| Transport Layer | TLS | DTLS |
| Network Layer | IPSec |  |

SIP Digest Authentication method uses five different headers for authentication namely: WWW-Authenticate, Authorization, Proxy-Authenticate, Proxy-Authorization and Authentication-Info. In this scheme, whenever any request from a SIP user agent needs to be authenticated, the recipient, for example a user agent, register or redirect server, challenges the originator by sending a 401 Unauthorized message with a WWW-Authenticate header. The header contains a nonce value and a realm in addition to some optional parameters. A nonce is a unique string generated each time the "401 Unauthorized" message is sent and the realm specifies the digest algorithm used for the challenge. After receiving this challenge, the client computes the response value using the nonce value, realm, username, shared secret with some optional parameters which is then included in the Authorization header in the new request message. By default, MD5 algorithm is used to compute this response value unless a different algorithm is specified in the realm parameter during the challenge. Fig. 2 illustrates the SIP Digest authentication mechanism.

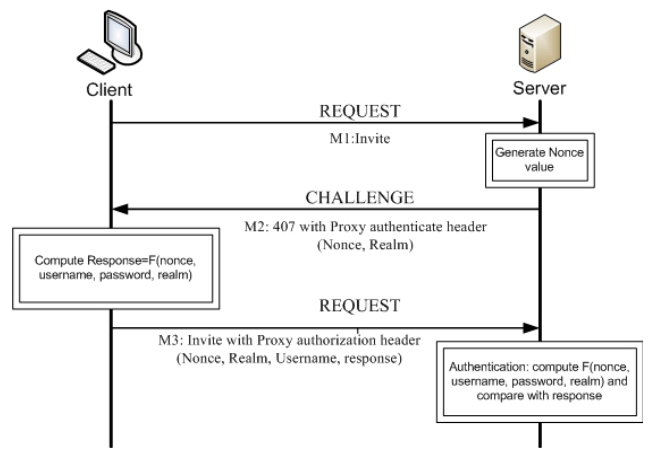

Fig. 2 SIP Digest authentication mechanism.

However, if the recipient is a proxy server, then a 407 unauthorized message with a Proxy-Authenticate header is sent instead of a 401 message to challenge the client. The client then includes a Proxy-Authorization header in reply to that challenge.

Upon receiving the re-issued request, the recipient should verify that the authenticated user is authorized to perform the requested action. If the authorization fails, the recipient may respond with another challenge.

## 2.2 S/MIME Authentication

SIP messages are capable of carrying the Multipurpose Internet Mail Extensions (MIME) bodies. Secure/Multipurpose Internet Mail Extensions (S/MIME) is used for providing the end to end confidentiality and integrity of the MIME content to some extent by replicating the header fields in the MIME part [8]. This scheme can also be used for authentication services by signing the replicated header fields to verify the identity of the sender. The S/MIME can be used to protect MIME payloads, such as the Session Description Protocol (SDP) data which is embedded on the application/sdp MIME body, contained in the SIP message by using the types multipart/signed or application/pkcs7-mim. S/MIME can be also used to protect the whole package through the creation of an S/MIME tunnel. Fig. 3 illustrates the SIP S/MIME authentication mechanism.

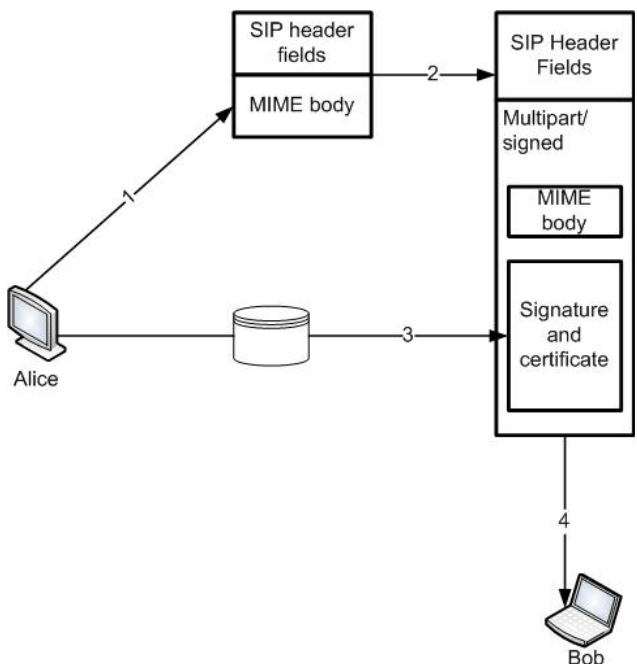

Fig. 3 SIP S/MIME authentication process.

In this process, SIP User Agent (UA) creates the SIP message and attaches a MIME body to it. Then the UA creates a multipart/signed S/MIME entity, that contains the MIME body and an application/pkcs7-signature

S/MIME entity. Afterwards, UA signs the MIME body by using its private key and includes the public key certificate in the CMS object of the application/pkcs7-signature S/MIME entity and then it sends the message. The receiving SIP UA takes the message apart and tries to find the included public key certificate in its keyring. If a public key certificate, in the keyring, has a subject that matches the From1 header field in the SIP message, then the receiving SIP UA should compare it with the received certificate. If there is a discrepancy between them, the SIP UA should notify the user and acquire the user's permission before continuing the session.

## 2.3 Proxy based Authentication

In the Proxy based authentication and encryption approach [9], a single certificate is used for a group of clients having the same role e.g. employees of a customer call center, to encrypt and sign the messages. In this scheme, an extension of the standard SIP proxy will be responsible for forwarding as well as signing and encrypting the messages. And for signing and encrypting the messages, a group based private key will be installed on this central proxy server. Generally, the validation of the signature and the decryption process will be done by the receiving terminal. However, these tasks can also be done by this proxy server on behalf of the clients and then the validated and decrypted messages will be forwarded to the receiving devices. Fig. 4 and Fig. 5 illustrate the call establishment and validation using proxy server.

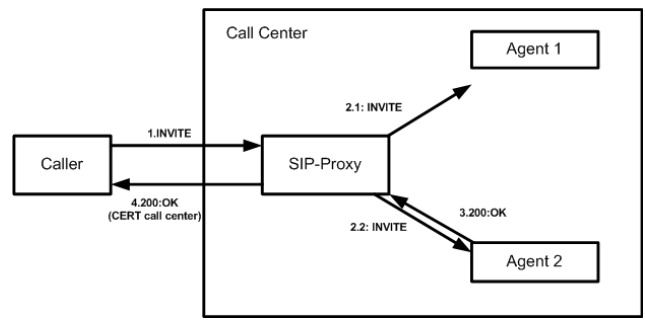

Fig. 4 Call establishment process with call center validated agent.

This is a suitable approach for specific scenarios such as the call center or support team of a company where it is convenient to have a group key, since it is easier for customers to manage just one certificate and, in those contexts, the person in the call center, for example, is signing the message as a representation of the call center answer and not as his answer (it may be desirable to provide anonymity to him).

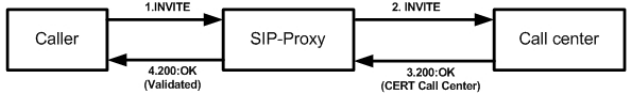

Fig. 5 Automatic proxy validation for call with call-center-validated agent.

Another benefit provided by having the encryption, decryption, and signing and signature verification on the proxy side is that all those costly operations no longer need to be done in the end-terminals behind the secure proxy, which is very convenient for devices with limited resources.

## 3. PKI less Authentication Mechanism and Key Agreement Protocols

Due to challenges on building up a global Public Key Infrastructure capable of supporting the possible strong authentication methods (the HTTP Digest Access Authentication by itself is not considered a strong authentication method, as pointed out in [7]), some PKI-less alternatives have been proposed to address the authentication issue in SIP.

### 3.1 Identity-based protocol

Identity-based (ID-based) protocol is based on the scheme of ID-based Cryptosystems, which was introduced with [10]. What Shamir proposes in his cryptographic scheme is that the public key of the user would be based (as a result of a predefined and global hash function for example) on one of his single identifier, such as his e-mail address, physical address, full name, ID number or even a combination of those. Then, the matching private key would be provided by a third party entity that would act as the Public Key Generator (PKG).

Once the public key corresponds to the user's identity, there is no need having the parties to exchange keys prior to the communication. Furthermore, it is possible to achieve non-repudiation and protection against the Man-in-the-Middle attack once the message is signed and it is easy to identify the owner of the signature based on the public key, which corresponds to the user identity. On the other hand, the security of the scheme relies on the trust in the PKG, on the security of the transmission of the private key from those PKGs to the user, on that the user will keep the private key securely, and, at last, on the security of the public key cryptographic algorithms used.

The public key scheme used for identity based cryptography must be so, that chosen the secret seed k, it is computably easy to generate a private key for a public key, but it is computably infeasible to trace the seed based on the public-private pairs generated, and, to attack one of the generated private keys with the possession of other generated pairs. This system ends up creating a key escrow (once the private keys are generated by a third party) and making it complex the process of key expiration and revocation, due to the identity nature of the public key. Although, one possible attempt to face the expiration problem is to have the identity key linked with its validity (the identity could be for example: "myself@mydomain.com + expiration-date"), and that the identity validity could be shortly limited, like to one day. Then, in order to revoke the key, it is just necessary that the PKG stops providing new private keys to the user. Adding the validity to the identity would also require to the user to get a new private key every day he needs to sign a message. Identity-based encryption proposed by Boneh [11] relies on the Weil pairing on elliptic curves where he pointed that the trust on the PKG in the identity-based scheme can be distributed among other sites through the use of threshold cryptography [12], and therefore reduce the risk severity of a compromised PKG and also raise the control over the key escrow.

The usage of the identity-based concept in SIP is presented by Ring et al. [13], where the id-based cryptosystem provides mutual authentication for the SIP protocol and, afterward, enable a key exchange agreement to generate the encryption key for the multimedia traffic. Concerning the authentication, it bases that the Proxy Server, would most likely represent the service provider, institution or company, and that it will act as the PKG.

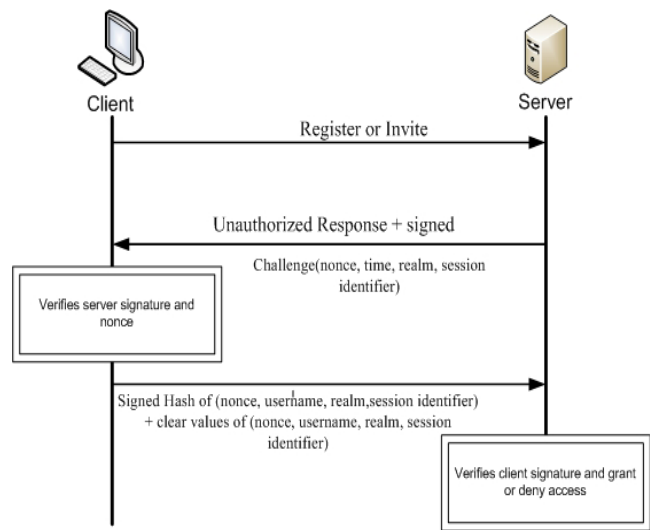

Fig. 6 ID-Based Authentication.

The authentication towards the Proxy is done through the usage of its public-private key and through the key pair of the client (besides a nonce created using the time stamp and server realm string (identity of the server) to strengthen the protocol against Man-in-the-Middle attacks and replay attacks). The authentication messages flow is described in the Fig. 6

The mutual authentication between both clients is achieved through the signing of the messages exchanged in the key exchange protocol, which works in the following manner:

- The initiator generates a temporary private key and sends his temporary public key signed by his permanent private key.
- The receptor verifies the signature against the sender's identification and also generates an ephemeral public pair, and reply with the temporary public key signed by his permanent private key.
- Then, both parties can generate the same temporary session key using their identities and public keys (both the ephemeral and the permanent one), with the bilinear pairing over elliptic curves.

J. Ring et al. argue that the usage of this authentication and key-exchange mechanism can be introduced in the SDP by just setting the "a" attribute defined with the values Additionally [13], they also argue that the usage of this authentication and key-exchange mechanism can be introduced in the SDP by just setting the "a" attribute defined with the values "a=idaka: base64 (ephemeral public key)" and "a=signature: base64 (signature)" [14].

The usage of the protocol would be limited in between SIP clients which have established a trusted relationship with a Trusted Authority and that have received a private key from it. There is a potential vulnerability when one of the parties is behind the PSTN network, since the authentication and key agreement would be done with a Media Gateway instead of the user.

## 3.2 Certificate-less Protocol

One alternative to the ID-based cryptography, which still does not rely on the PKI, is the certificate-less public key cryptography (CL-PKC) [15]. This approach still makes the use of a PKG, but the PKG only generates a component of the user's private key (still based on his identity), while the second component is generated by the user. The trust relied on the PKG becomes similar to the one given to a Certification Authority (CA), that it won't issue fake keys. Since the PKG does not have the user's complete private key, it can't decrypt or sign his messages, as a consequence, the key escrow is eliminated.

- The steps necessary to set-up the certificate-less key pair are described below. Some of the steps do not

need to be performed in the same sequence they are presented.

- The PKG generates his master-key besides the public parameters for the public key cryptosystem using bilinear Diffie-Hellman Pairing.
- The PKG generates the partial private key for the user, using the user's identity as an input to the key generation algorithm.
- The user chooses a secret value for him.
- The user generates his public key based on the public parameters and his secret value.
- The user generates his final private key, using the chosen secret and the partial private key provided by the PKG as an input to the key generation algorithm, which could also use his public key and identity in order to bind his key.

It is worth to mention that Al-Riyami et al. propose the possibility of binding the identity parameter of the key to the public key, and therefore allowing the user to create only one public key for the same partial private key, besides, making it noticeable in case the PKG try to replace a public key of the user by one it knows the whole secret key.

In order that the message is encrypted or the signature is verified, it is necessary that the users exchange their public keys or make them available at a public directory. Nevertheless, there is no need for any Authority to certify the key. The transfer and the checking of certificates is avoided, and consequently, their bandwidth and computational (processor and memory usage) overhead.

The usage of CL-PKC in SIP towards the accomplishment of mutual authentication and key agreement, in a very similar way to the one presented in the previous item, is presented at [16]. The protocol presented by Wang et al. [16] is different from the proposal of Al-Riyami et al. [15], as the mutual authentication can be done in between clients independently of their security domains once their identities are bound to their public keys. The key escrow is eliminated since the PKGs no longer posses the user's private keys.

## 4. Analysis of Different Authentication Methods

In this section we have highlighted the limitations of different authentication methods.

## 4.1 SIP Digest Authentication

SIP Digest Authentication resolves some of the deficiencies of HTTP basic authentication approach by transmitting a MD5 or SHA-1 digest of the communicating parties' credentials. Besides that, the use of the unique nonce value to challenge the sender prevents the reply attack.

However, there are also some limitations of Digest authentication. One major problem of Digest authentication is that it requires pre-existing secure association for password distribution. As a result this scheme cannot ensure the security in a non-secure association such as proxy-to-proxy authentication across different domains of internet. The authentication scheme in such case has to rely on TLS or IPSec. The digest authentication is also vulnerable against chosen plaintext attack. If short or weak passwords are used then the attacker can intercepts many responses from different nonces, and then tries out the password with brute force. Another drawback of SIP Digest authentication scheme is that it doesn't include any mechanism for proxy to User Agent Server (UAS) authentication. As a result, the authentication of the last hop in a SIP call routing is not possible using SIP Digest Authentication unless any lower layer security protocol is used. SIP specification recommends the use of Transport Layer security (TLS) or Datagram TLS (DTLS) [17] protocol to provide adequate level of protection to against the attacks. In addition to these methods, an extension of SIP digest authentication scheme is proposed to solve proxy-to-UAS authentication problem [18]. In this extension, some new headers are defined for the authentication such as:

- 492 proxies unauthorized: used by UAS to challenge the proxy server
- UAS-Authenticate Response Header: response to the challenge containing the credential of the proxy server.
- UAS-Authorization Request Header: used by the proxy to send a request for the credentials of the UAS for mutual authentication.
- UAS-Authentication info: used by UAS to authenticate itself to proxy server.

The steps of the proposed authentication procedure are given in the Fig. 7.

## 4.2 S/MIME

SIP RFC recommends the replication of all header fields inside a MIME part, which exposes some problems. First, the SIP header fields might get altered by the intermediate SIP entities which make it difficult for the recipient to identify the legal or malicious changes in headers. Second, S/MIME solution also mandates the deployment of a global S/MIME Public Key Infrastructure (PKI). Otherwise, the exchanged public keys would be self-signed, which makes the initial key exchange susceptible to man-in-the-middle attacks [19]. Third, SIP messages can be large by their size, which causes overhead for processing and transporting of the messages.

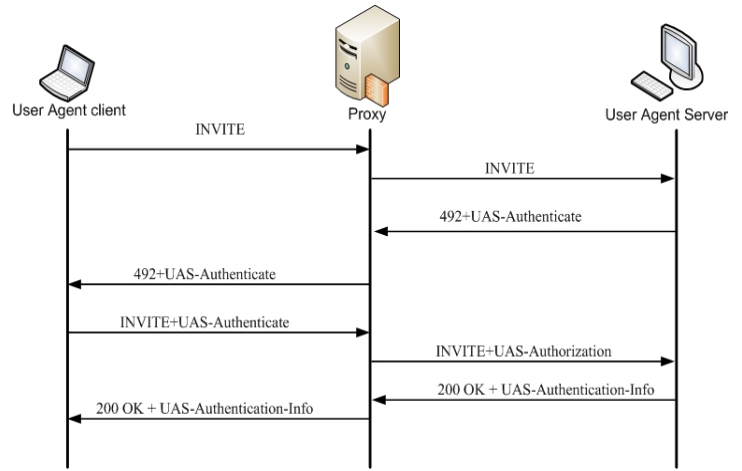

Fig. 7 Proxy to UAS authentication using extended SIP Digest authentication.

The possible solution of this problem has been discussed by Peterson [20], where he has proposed a new Authenticated Identity Body (AIB) to deliver the authenticated identity of the call parties. In SIP, AIB is basically a MIME body containing the authenticated identities.

## 4.3 ID-based Cryptographic Authentication

The ID-based protocol eliminates the need of PKI and makes it unnecessary the transmissions of the public keys, but at the expense of the need of relying in the security of a PKG, and the key escrow facility. The key escrow, depending on the context, can be seen as an advantage since it would be realistic to foresee that some countries may require the possibility of wire-tapping the communication between the users. It also does not allow the peer-to-peer usage whenever one of the users does not have his ID-based key pair, making it necessary that both of the parties have a trust relation with a PKG.

## 4.4 Certificate-less mutual authentication

This approach also eliminates the need of the PKI, although it is necessary for the parties to exchange their public keys.

| Authentication Methods:<br>PSK: Pre shared key<br>PKI: Public key infrastructure<br>ID: Identity based cryptography | Authentication | Data Integrity | Confidentiality | |
|---|---|---|---|---|
| HTTP Basic Authentication | PSK | - | - | Insecure transmission of user's credentials. Deprecated by SIPv2 |
| HTTP Digest Authentication | PSK | - | - | Challenge/response exchange based ON md5 hash of password |
| Secure Mime (S/MIME) | PKI | √ | √ | For encryption the public key of the recipient user agent must be known |
| DTLS | PKI | √ | √ | |
| Proxy Based authentication | PKI | √ | √ | |
| Id based authentication | ID | √ | √ | |
| Certificate less authentication | | √ | √ | |

Fig. 8 Summary of different authentication mechanisms.

The advantage over the ID-based system is that the trust on the PKG is drastically reduced and the key escrow is eliminated. However, there is no process of revoking the keys. The Fig. 8 summarizes key features o different authentication mechanisms.

## 4. Conclusions

After analyzing the security aspects of the available methods, we can clearly see a move towards alternatives that do rely on neither the pre-shared keys nor a PKI. It is noticeable that the different approaches better suit different scenarios and they most-likely have focused on those scenarios. For a context where key escrow is necessary, the ID-Based solution should fit very well, while at the same time, proxy based authentication would be very suitable for a support or call center. From our analysis, we can see the appearance of a few new proposed Identity based authentication protocols that might lead to more satisfactory solution.

## References

[1] J. Rosenberg, H. Schulzrinne, G. Camarillo, A. Johnston, J. Peterson, R. Sparks, M. Handley and E. Schooler, "SIP: Session Initiation Protocol," RFC 3261, June 1999. [Online]. Available: http://www.faqs.org/rfcs/rfc3261.html

[2] B. Campbell, J. Rosenberg, H. Schulzrinne, C. Huitema and D. Gurle, "Session Initiation Protocol (SIP) Extension for Instant Messaging," RFC 3428, December 2002. [Online]. Available: http://www.faqs.org/rfcs/rfc3428.html

[3] A. Singh and A. Acharya, "Multiplayer networked gaming with the session initiation protocol," in Computer Networks: The International Journal of Computer and Telecommunications Networking. Amsterdam, The Netherlands: Elsevier Science, 2005, vol. 49, pp. 212–223.

[4] "ITU-T Recommendation H.323: Packet-based Multimedia Communications Systems," International Telecommunications Union, Jun. 2006.

[5] IP Multimedia (IM) Subsystem - Stage 2 (Release 8), 3GPP TS 23.228 V8.6.0, 3rd Generation Partnership Project Std., September 2001. [Online]. Available: http://www.3gpp.org

[6] A. Johnston, S. Donovan, R. Sparks, C. Cunningham and K. Summers, "Session Initiation Protocol (SIP) Basic Call Flow Examples," RFC 3665, December 2003. [Online]. Available: http://www.faqs.org/rfcs/rfc3665.html

[7] J. Franks, P. Hallam-Baker, J. Hostetler, S. Lawrence, P. Leach, A. Luotonen, L. Stewart, "HTTP Authentication: Basic and Digest Access Authentication," RFC 2617, June 1999. [Online]. Available: http://www.faqs.org/rfcs/rfc2617.html

[8] B. Ramsdell, "S/MIME Version 3 Message Specification," RFC 2633, June 1999. [Online]. Available: http://www.faqs.org/rfcs/rfc2633.html

[9] J. Arkko, E. Carrara, F. Lindholm, M. Naslund and K. Norrman, "MIKEY: Multimedia Internet KEYing," RFC 3830, August 2004. [Online]. Available: http://www.faqs.org/rfcs/rfc3830.html

[10] A. Shamir, Identity-Based Cryptosystems and Signature Schemes, ser. Lecture Notes In Computer Science. London, UK: Springer-Verlag, 1984, vol. 196.

[11] D. Boneh, M. Franklin, Identity Based Encryption from the Weil Pairing, 2003, vol. 32.

[12] P. Gemmell, "An introduction to threshold cryptography," CryptoBytes, a technical newsletter of RSA Laboratories, vol. 2, 1997.

[13] J. Ring, K.-K. R. Choo, E. Foo, and M. Looi, "A new authentication mechanism and key agreement protocol for SIP using identity-based cryptography," in Proc. AusCERT Asia Pacific Information Technology Security Conference 2006, Gold Coast, Australia, May 2006, pp. 61–72.

[14] M. Handley and V. Jacobson, "SDP: Session Description Protocol," RFC 2327, April 1998. [Online]. Available: http://www.faqs.org/rfcs/rfc2327.html

[15] S. Al-Riyami and K. Paterson, Certificateless public key cryptography, ser. Lecture Notes In Computer Science. Heidelberg, Germany: Springer-Verlag, 2003, vol. 2894.

[16] F. Wang and Y. Zhang, "A New Provably Secure Authentication and Key Agreement Mechanism for SIP Using Certificateless Public-key Cryptography," in Computer Communications: The International Journal for the Computer and Telecommunications Industry. Amsterdam, The Netherlands: Elsevier Science, 2008, vol. 31, no. 10, pp. 2142–2149.

[17] E. Rescorla, N. Modadugu, "Datagram Transport Layer Security," RFC 4347, April 2006. [Online]. Available: http://www.faqs.org/rfcs/rfc4347.html

[18] Q. Qiu, "Study of digest authentication for Session Initiation protocol (SIP)," M.Sc. Thesis, University of Ottawa, Ottawa, Canada, December 2003. [Online]. Available: http://www.site.uottawa.ca/~bob/gradstudents/DigestAuthenticationReport.pdf

[19] C. Chang, Y. Lu, A. Pang, and T. Kuo, Design and Implementation of SIP Security, ser. Lecture Notes In Computer Science. Heidelberg, Germany: Springer-Verlag, 2005, vol. 3391.

[20] J. Peterson, "Session Initiation Protocol (SIP) Authenticated Identity Body (AIB) Format," RFC 3893, September 2004. [Online]. Available: http://www.faqs.org/rfcs/rfc3893.html

**Abdullah Al Hasib** received dual M.Sc. degree in Security and Mobile Computing from Helsinki University of Technology (TKK), Finland and Royal Institute of Technology (KTH), Sweden in 2009. He obtained his B.Sc. degree from Islamic University of Technology (IUT), Bangladesh in 2005. He got Erasmus Mundus scholarship from European Union for his M.Sc studies. Presently, he is a faculty member at Computer Science and Information Technology department in IUT, Bangladesh. His research interest includes cryptographic protocols, wireless network security and mobility management.

**Abdullah Azfar** is doing his MS in Erasmus Mundus NordSecMob program specialized in Security and Mobile Computing in Norwegian University of Science and Technology (NTNU), Norway and Royal Institute of Technology (KTH), Sweden. He received his BSc degree in Computer Science and Information Technology from Islamic University of Technology (IUT), Gazipur, Bangladesh in 2005. He served as a lecturer in the Islamic University of Technology during the period March 2006 – July 2008. He also served as a lecturer in Prime University, Dhaka, Bangladesh during the period October 2005 – February 2006. He received the Erasmus Mundus scholarship from the European Union for his MS studies and OIC (Organization of the Islamic Conference) scholarship for three years during his BSc studies. His research interest is mainly focused on Information Systems Security. At present he is working with security issues in VoIP.

**Md. Sarwar Morshed** received his M.Sc. degree in Information Technology with major in Communication Engineering from Tampere University of Technology in 2009. He received his B.Sc. degree from Islamic University of Technology in 2005. He is currently a Lecturer in Islamic University of Technology. His research interests include wireless communication, internetworking and security in transmission.